\pdfoutput=1
\def\year{2019}\relax

\documentclass[letterpaper]{article} %DO NOT CHANGE THIS
\usepackage{aaai}  %Required
\usepackage{times}  %Required
\usepackage{helvet}  %Required
\usepackage{courier}  %Required
\usepackage{url}  %Required
\usepackage{graphicx}  %Required
\usepackage{amsmath,amssymb,amsthm}
\usepackage{url}

\begin{document}
 
\setlength{\abovedisplayskip}{4pt}
\setlength{\belowdisplayskip}{4pt}
\title{What Should I Learn First: Introducing LectureBank \\for NLP Education and Prerequisite Chain Learning}
% \title{What should I learn first: Introducing LectureBank \\ for prerequisite chain learning}

\author{{\large Irene Li \quad\quad Alexander R. Fabbri \quad\quad Robert R. Tung \quad \quad Dragomir R. Radev} \\
Department of Computer Science, Yale University\\
\scalebox{0.85}[0.9]{{\tt \{irene.li,alexander.fabbri,robert.tung,dragomir.radev\}@yale.edu}}
% \scalebox{0.85}[0.9]{}
}

\maketitle
\begin{abstract}
Recent years have witnessed the rising popularity of Natural Language Processing (NLP) and related fields such as Artificial Intelligence (AI) and Machine Learning (ML). Many online courses and resources are available even for those without a strong background in the field. Often the student is curious about a specific topic but does not quite know where to begin studying. To answer the question of \lq\lq what should one learn first,\rq\rq we apply an embedding-based method to learn prerequisite relations for course concepts in the domain of NLP. We introduce LectureBank, a dataset containing 1,352 English lecture files collected from university courses which are each classified according to an existing taxonomy as well as 208 manually-labeled prerequisite relation topics, which is publicly available \footnote{\url{https://github.com/Yale-LILY/LectureBank}}. The dataset will be useful for educational purposes such as lecture preparation and organization as well as applications such as reading list generation. Additionally, we experiment with neural graph-based networks and non-neural classifiers to learn these prerequisite relations from our dataset. 
\end{abstract}

\section{Introduction}

\begin{figure*}[!htb]
  \includegraphics[width=7in]{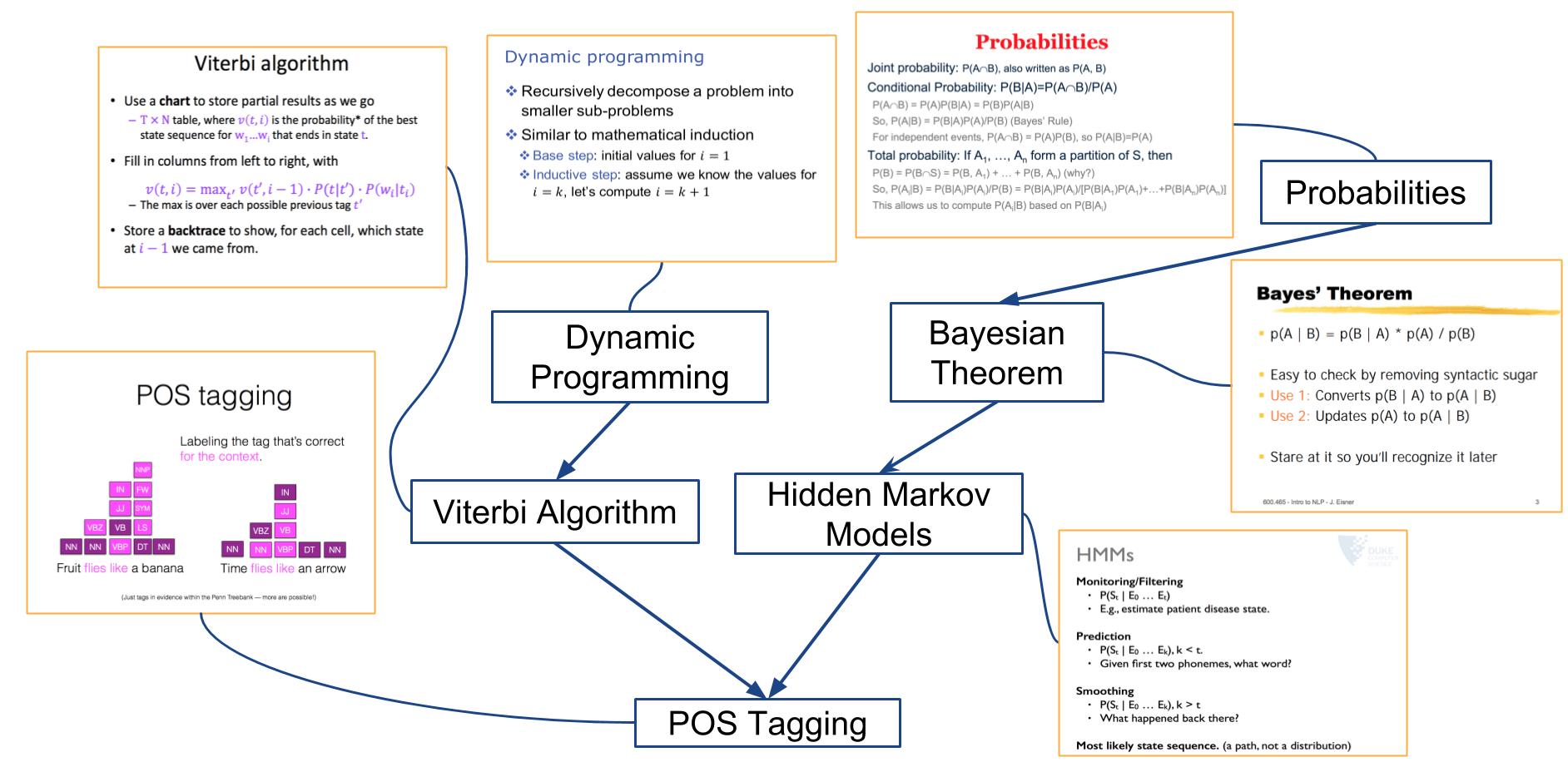}
  \caption[Caption for Prerequisites]{An example of prerequisite relations from lecture slides depicted as a directed graph. The direction of an edge is from a prerequisite of a concept to the concept itself. For example, \textit{Hidden Markov Models} is a prerequisite of \textit{POS Tagging}. We illustrate each concept with a slide on that topic, selected from our corpus. 
  The references for the slides starting from POS Tagging and moving clockwise are: 
  \cite{ppt1}, \cite{ppt2}, \cite{ppt3}, \cite{ppt4}, \cite{ppt5} and \cite{ppt6}. }
  \label{fig:example}
\end{figure*}

As more and more online courses and resources are becoming accessible to the general public, researching advanced topics has become more feasible. A large amount of educational material is available online, although it is not structured in an organized way. \cite{margolis1999concepts} suggested that \textit{concepts} are one of the most fundamental constructs and that having an order for learning and organizing them is essential for acquiring new knowledge. To be able to capture the concept organization and dependencies for NLP, we study the problem of building concept prerequisite chains. We treat each concept as a single vertex, and learn the dependencies to ultimately build a concept graph as in \cite{gordon2016modeling}.  We define a \textit{prerequisite} to be the directed dependency between two vertices. Once prerequisite relations among concepts are learned, these relations can be used for downstream tasks and applications such as generating reading lists for readers based on a query as well as curriculum planning \cite{gordon2017structured}.
\par

% example with the diagram
Imagine the scenario in Figure \ref{fig:example} in which a student has some basic knowledge of NLP but wants to learn a specific new concept such as \textit{POS tagging.} In order to fully understand this concept, he or she should have an understanding of prerequisite concepts such as \textit{Viterbi Algorithm} and \textit{Markov Models}, as well as the prerequisites for these concepts: \textit{Dynamic Programming}  \textit{Bayes Theorem} and \textit{Probabilities.}  Although many search engines can provide relevant documents or learning resources, most of the results are based on relevancy and semantics while few have the ability to provide a reasonable list based on a path to learning a new concept.  Additionally, we might want to recommend the corresponding learning materials such as lecture files for each of the concepts. Thus, for educational purposes, we aim to learn the prerequisite relations of each concept pair and eventually apply them in a search engine. Given a query concept word or phrase, the search engine would provide the study materials corresponding to the prerequisites of the query concept.
  \par
Recent work has focused on extracting and learning concept dependencies from scientific corpora including the ACL Anthology \cite{Bird:08} as well as from online courses \cite{gordon2016modeling,pan2017prerequisite,liu2016learning}. On the other hand, we are more interested in learning materials such as blogs, surveys and papers, and in this paper, we focus on lecture files, which are usually well-organized and contain a focused topic. We manually collected lecture files by inspection of the file contents and annotated each lecture file according to a taxonomy of 305 topics. We used our LectureBank corpus and a recently published corpus TutorialBank \cite{fabbri2018tutorialbank} as training data. We annotated prerequisite relations for each concept pair from a list of 208 concepts provided in \cite{fabbri2018tutorialbank}. These 208 concepts differ in granularity and scope from the 305 taxonomy topics; they consist of topics for which one could conceivably write a survey and are thus neither too fine-grained or large in scope. Similarly to \cite{pan2017prerequisite}, we focus on learning embedded representations of the concepts. We test the effectiveness of standard classifiers as well as the recently introduced neural link-prediction approaches of Variational\cite{kipf2016variational} and vanilla Graph Autoencoders  \cite{schlichtkrull2018modeling} to discover prerequisite relations.

Our main contributions are the following. First, we introduce our LectureBank dataset with 1,352 English lecture files (51,939 slides) classified according to an existing taxonomy. The dataset can be used directly as study material mainly in the fields of NLP or ML, as it covers high-quality university lectures suitable for entry-level researchers or NLP engineers working at the internet or social media companies. The corpus can also be used for topic modeling of scientific topics in addition to the prerequisite chains learning task. Additional details on the dataset can be found in Section 3. Second, we compare novel graph-based deep learning models, which have shown promise in the task of link prediction, with standard classification methods and demonstrate the importance of oversampling in this task.

\section{Related Work}
In this section, we briefly describe related work on prerequisite chain learning using concept graphs as well as recent developments in neural graph-based methods.

\subsection{Concept Graphs}
%add cmu paper
Previous work collected online courses including Computer Science,  Statistics and Mathematics from Massachusetts Institute of Technology
, California Institute of Technology, Princeton University and Carnegie Mellon University and proposed an approach for inference within and across course-level and concept-level directed graphs \cite{liu2016learning}. We focused on detailed concept-level mainly in the NLP domain. 
\cite{gordon2016modeling} introduced two methods for discovering concept dependency relations automatically from a text corpus: a cross-entropy approach and an information-flow approach. They tested the methods on the ACL Anthology corpus \cite{Bird:08}. Concepts were determined using LDA topic modeling \cite{blei2003latent}, and prerequisite relations are notably found in an unsupervised setting.  \cite{pan2017course} proposed a representation learning-based method to learn the concepts and a set of novel features to identify prerequisite relations.  However, these methods benefit from the help of Wikipedia to perform entity extraction and improve entity representations. However, a Wikipedia page may not always be available for a concept, and we focus on obtaining concept representations solely from our corpus. 

Other work has focused on generating prerequisite relations among concepts on a Massive Open Online Courses (MOOCs) corpus \cite{pan2017prerequisite}. They proposed seven types of features covering course concept semantics, course video context and course structure. They cast the prerequisite chain problem as a binary classification problem by evaluating the prerequisite relationship between each concept pair and applying four different binary classifiers. While they constructed three datasets to evaluate their methods, the coverage of the datasets is relatively small, as only up to 244 topics and three domains (Machine Learning, Data Structure and Algorithms and Calculus) are considered.  In this paper, we expand the range of domains and the number of courses. 

% Moved
Additionally, a manually-collected and categorized corpus of about 6,300  resources on NLP and the related fields of ML, AI, and IR was recently introduced by \cite{fabbri2018tutorialbank}. This corpus was released with a list of 208 topics for which prerequisite relationships were annotated, making it a complementary dataset to ours. However, only a single annotator annotated each relation and prerequisite annotations were ternary versus our binary classification. As there have not been any experiments on learning prerequisite chains using this corpus, we choose to take advantage of both TutorialBank and LectureBank to learn prerequisite relations. Besides the 208 topics for prerequisite relationships, they also proposed a university-level NLP course syllabus-like taxonomy of 305 topics, which the surveys, tutorials and resources other than scientific papers are classified into. These 305 topics can be coarse-grained, such as \textit{Natural Language Processing}, \textit{Artificial Intelligence}. Also, there are redundant topics such as \textit{Classification and kNN 1} and \textit{Classification and kNN 2}. In comparison, the 208 topics are more fine-grained and suitable for prerequisite chain learning, although there are some overlap topics with the 305 topics. Of note, as in the TutorialBank dataset, deep learning topics are a major focus as both datasets consist of work from the last few years, and an abundance of tutorials and resources on deep learning have been published in this time, thus explaining this bias. 
% The readers might have noticed that deep learning topics are more focused, as in the TutorialBank, resources are predominantly from the last few years and deep learning topics haven been an abundance of tutorials and resources published about deep learning, so a bias towards deep learning topics can be seen.
%  The 305 topics form a tree, but the relationship among the topics is that of encompassing topics rather than being a prerequisite of a topic, thus the need for the prerequisite annotations. 

\subsection{Graph Convolutional Networks}
Using neural networks on structured data structures such as graphs is a difficult problem. Graph Convolutional Networks (GCNs) aim to tackle this task by drawing inspiration from spectral graph theory. \cite{Defferrard16} design fast localized convolutional filters on graphs for an image-processing task while \cite{kipf:16} apply GCNs to a number of semi-supervised graph-based classification tasks, reporting faster training times and better predictive accuracy. More recently GCNs have been applied to problems such as machine translation \cite{Bastings17} and summarization \cite{yasunaga2017graph}. GCNs have also been applied to the task of link prediction or predicting relations among entities.
For this task, some, but not all, of the relations among entities are given during training, and the goal is to predict additional unobserved relations during testing. Finding prerequisite relations can be viewed as link prediction, where the vertices are concepts, and the edges are the prerequisite relations, or lack thereof. 
\par
The work from \cite{kipf2016variational} introduced a non-probabilistic Graph Autoencoder (GAE) as well as the Variational Graph Autoencoder (VGAE). These models build upon work on autoencoders and variational autoencoders for unsupervised learning on graph-structured data for tasks such as predicting links in a citation network. These models combine a GCN encoder with inner product decoders and latent variables in the case of VGAE.  \cite{schlichtkrull2018modeling} extend GCNs and variational graph autoencoders to model large-scale relational data for the tasks of link prediction and entity classification, and thus we examine the applicability of these models for the prerequisite chains learning task.  

\section{LectureBank Dataset}

In this section, we introduce the LectureBank dataset, analysis, statistics, annotations,  and then compare it with other similar datasets. 

%TODO: add Taxonomy (!) (explain the topics,what are we focusing, etc); add # of pages, ect.

% fisheye view
 
%  courses sampling, also a few courses for smaller topics (ai, ir) 

% \subsection{Data Collection, Annotation and Presentation}
\subsection{Data Collection and Presentation}
% Overview
We collected online lecture files from 60 courses covering 5 different domains, including NLP, ML, AI, deep learning (DL) and information retrieval (IR). For copyright reasons, we are releasing only the links to the individual lectures. 

In order to make the course slides more accessible to the user, we have indexed all slide lectures and created a search engine which allows the user to browse courses and slides according to queries. As related material on a subject can come from a variety of courses, we believe gathering them into a single search engine dramatically reduces the amount of time a student will spend searching for relevant material. The search engine will be made publicly available. 

% TODO make analysis its own section
\subsection{Dataset Analysis and Statistics}
Our LectureBank dataset focuses on courses from 5 domains, and detailed statistics can be found in Table \ref{tab:courses}. The table reports the number of courses, lecture files, pages tokens, the average tokens per lecture file and the average tokens per page. For preprocessing, we used the PDFMiner\footnote{https://pypi.python.org/pypi/pdfminer3k/} python package to extract the texts from the PDF files, and python-pptx\footnote{https://python-pptx.readthedocs.io/en/latest/\#} to extract Powerpoint (PPT) presentations. If a course provided both PDF and PPT versions, we kept the PDF files and removed the PPT files.

\begin{table*}[t]
\small
    \centering
\begin{tabular}{ |c c c c c c c| }
\hline
 Domain & \#courses & \#lectures  & \#slides & \#tokens  & \#tokens/lecture  & \#tokens/slide \\ 
 \hline
NLP	&	35	&	764	&	29,661	&	1,570,578	&	2,055.73	&	52.95	\\	\hline 
ML	&	12	&	260	&	10,720	&	866,728	&	3,333.57	&	80.85	\\	\hline
AI	&	5	&	101	&	4,911	&	265,460	&	2,628.32	&	54.05	\\	\hline
DL	&	4	&	148	&	3,270	&	582,502	&	3,935.82	&	178.14	\\	\hline
IR	&	4	&	79	&	3,377	&	157,808	&	1,997.57	&	46.73	\\	\hline

\textbf{Overall}	&	60	&	1352	&	51,939	&	3,443,076	&	2,546.65	&	66.29	\\	\hline
\end{tabular}
    \caption{LectureBank Dataset Statistics: within each domain, we have a certain number of courses; each course consists of lectures files; each lecture file has multiple individual slides. The contrasting number of tokens per slide for DL is a result of the small sample size and the courses chosen.}
    \label{tab:courses}
\end{table*}

\subsection{Additional Annotation}
% TODO, cite CMU paper
\textbf{Prerequisite Annotation} In addition to using our own dataset for prerequisite chain learning, we make use of a recently introduced corpus of resources on topics related to NLP. \cite{fabbri2018tutorialbank} introduced a set of 208 topics on NLP and related fields and annotated all of the concept pairs as one of the following: not a prerequisite, somewhat a prerequisite or a true prerequisite. While the annotations emphasize the breadth of pairs annotated, we think certain dependency relations might not be clear to annotate, such as "long dependencies" (if A is a prerequisite of B and B is a prerequisite of C, should we count A as a prerequisite of C or not). We aim to improve classification agreement by making the criteria more precise and having additional annotators annotate the topics. Each of our annotators annotated each prerequisite relation, and we report high inter-annotator agreement. Our annotators consist of two PhD students working on NLP. We obtained a Cohen's kappa \cite{cohen1960coefficient} of 0.7 which according to \cite{Landis:1977} is considered substantial agreement. We asked the annotators the following question for each topic pair (A, B): is A a prerequisite of B; i.e.,  do you think learning the concept A will help one to learn the concept B? Even if the distance between two concepts in a potential concept graph is large, if one topic is typically learned before the other in a university course and is helpful in building up knowledge for learning another concept, we considered this earlier concept a prerequisite of the other. Thus, while the criteria may be subjective, we direct the annotators to refer to standard university curricula for unclear prerequisite pairs. Only binary \textit{yes} (positive) or \textit{no} (negative)  answers are possible. This is in contrast to the ternary classification of \cite{fabbri2018tutorialbank}, who report a  kappa score of .3 on the same prerequisite pairs.  We chose binary annotation over the ternary annotation of \cite{fabbri2018tutorialbank} as this is the same setup as in related work such as \cite{pan2017prerequisite}. Additionally, the choice of binary as opposed to ternary classification pertains to the precision and recall trade-off which we discuss in the results section below. We decided that concepts which are "somewhat prerequisites" as in \cite{fabbri2018tutorialbank} should be labelled as prerequisites so that they are not missed as potential missing areas of knowledge. 

We took the intersection of the two annotators' annotations,  which resulted in a labeled directed concept graph with 208 concept vertices and 921 edges. If concept A is a prerequisite of concept B, the edge direction goes from concept vertex A to concept vertex B.  We also observed some cycles between a pair of vertices within the concept graph. We found 12 such pairs in our labeled concept graph. These pairs consist of very closely related topics such as \textit{Domain Adaptation} and \textit{Transfer Learning} and \textit{LDA} and \textit{Topic Modeling}, suggesting that in the future we may combine these pairs into a single concept. There are 7 independent topics which have no prerequisite relationships with the rest of the topics. They are: \textit{Morphological Disambiguation, Weakly-supervised learning, Multi-task Learning, ImageNet, Human-robot interaction, Game playing in AI, data structures and algorithms}. The topics were proposed by \cite{fabbri2018tutorialbank}, and so we chose to keep all the topics in our experiments.

% For example, the pair \textit{BLEU} and \textit{ROUGE} are very similar, when doing the annotating, we consider learning each one will be helpful to learn the other one. Also the pair \textit{Domain Adaptation} and \textit{Transfer Learning} are very similar in our corpus  understanding.
% showing in Table \ref{tab:cycle_pairs}, suggesting the potentials to combine each concept
%  replace naive bayes with bayes theorem 
% \begin{table}[b]
% \tiny
%     \centering
% \begin{tabular}{|c c| }
% \hline
% Neural Machine Translation & BLEU \\ \hline
% Neural Machine Translation & ROUGE \\ \hline
% IBM Translation Models & BLEU \\ \hline
% IBM Translation Models & ROUGE \\ \hline
% BLEU & ROUGE \\ \hline
% Knowledge Graph & Knowledge representation \\ \hline
% Domain Adaptation & Transfer Learning \\ \hline
% Latent Dirichlet Allocation & Topic Modeling \\ \hline
% Dimensionality Reduction & Singular-value Decomposition \\ \hline
% First-order Logic & Knowledge Representation \\ \hline
% Knowledge Representation & Expert Systems \\ \hline
% Backpropagation Through Time & Artificial Neural Networks \\ \hline

% \end{tabular}
%     \caption{Twelve Cycle Pairs: each row contains a pair of concepts. }
%     \label{tab:cycle_pairs}
% \end{table}

We also list the concept vertices that have the largest in-degree and out-degree in Table \ref{tab:degree}. In-degree illustrates that the concept vertex has many prerequisite concepts; out-degree illustrates that the concept vertex is a prerequisite to many other concepts. The concepts with large in-degree are advanced concepts which require much background knowledge in order to be learned well, while the list of concepts with large out-degree are more fundamental concepts.  We also observed the longest path in the constructed concept graph, which consists of 14 concepts in the path: \textit{Matrix Multiplication, Differential Calculus, Backpropagation, Backpropagation Through Time, Artificial Neural Network, Word Embeddings, Word2Vec, Seq2Seq, Neural Machine Translation, BLEU, IBM Translation Models, ROUGE, Automatic Summarization, Scientific Article Summarization}.

\begin{table*}[h]
\small
    \centering
\begin{tabular}{|c|c||c|c| }
\hline
Most in-degree Concept Vertices & Count & Most out-degree Concept Vertices & Count\\ 
\hline
\hline
Neural Machine Translation	&	19	&	Data Structures and Algorithms		&	106	\\	\hline
Variational Autoencoders	&	15	&	Probabilities		&	105	\\	\hline
Stack LSTM	&	13	&	Linear Algebra		&	98	\\	\hline
Seq2seq	Models &	13	&	Matrix Multiplication		&	72	\\	\hline
Highway Networks	&	12	&	Bayes Theorem		&	59	\\	\hline
DQN	&	12	&	Conditional Probability		&	58	\\	\hline
Bidirectional Recurrent Neural Networks	&	11	&	Differential Calculus		&	21	\\	\hline
Convolutional Neural Networks	&	11	&	Activation Functions		&	20	\\	\hline
Multilingual Word Embeddings	&	11	&	Loss Function		&	19	\\	\hline
Capsule Networks	&	11	&	Entropy		&	17	\\	\hline
Topic Modeling	&	10	&	Data Preprocessing		&	17	\\	\hline
Neural Turing Machine	&	10	&	Backpropagation		&	17	\\	\hline
Recursive Neural Networks	&	10	&	Artificial Neural Networks		&	16	\\	\hline
Attention Models	&	10	&	Backpropagation Through Time		&	14	\\	\hline
Generative Adversarial Networks	&	10	&	Information Theory		&	13	\\	\hline

\end{tabular}
    \caption{Concept vertices from our annotated concept graph with the largest in-degree and out-degree}
    \label{tab:degree}
\end{table*}

% Need to clarify 305 topics with 208 topics
\textbf{Classification} We used the TutorialBank taxonomy from \cite{fabbri2018tutorialbank}, which contains 305 topics of varying granularity. Based on a university-level NLP course syllabus, the TutorialBank taxonomy was then expanded to other topics from other courses from IR, AI, ML and DL. The 305 taxonomy topics cover a wide range of topics in NLP area, and we only use these topics during our manual labeling of our lecture dataset as an additional annotation work. 
We manually classified all 1,352 LectureBank lecture files into the TutorialBank taxonomy. In Table \ref{tab:stat_taxonomy} we show the top 10 more frequent TutorialBank taxonomy labels for our corpus classification.

Another taxonomy which was considered for our classification was the 2012 ACM Computing Classification System (CCS) \footnote{https://www.acm.org/publications/class-2012}, a poly-hierarchical ontology that can be utilized in semantic web applications. The system is hierarchically structured into four levels; \textit{Machine Translation}, for example, can be found in the following branch: \textit{Computing Methodologies} $\rightarrow$ \textit{Artificial Intelligence} $\rightarrow$ \textit{Natural Language Processing} $\rightarrow$ \textit{Machine Translation}. The \textit{Artificial Intelligence} directory has 8 subcategories including \textit{Natural Language processing}, and 8 subcategories under the \textit{Natural Language Processing} directory. However, rather than focusing on the larger scope of computing or even AI in general, our main focus in NLP. Compared with CCS, the TutorialBank taxonomy covers detailed categorization, focusing on NLP and related fields insofar as they related to NLP, making it very suitable for our desired classification. 
\begin{table}

\small
    \centering
% \begin{center}
% \begin{tabular}{ |p{4.5cm}||p{1cm}| }
\begin{tabular}{|c|c|}
\hline
Topic   & Count \\ \hline \hline
Introduction to Neural Networks & 92    \\ \hline
Machine Learning Resources                                                                   & 56    \\ \hline
Information Retrieval                                                                        & 31    \\ \hline
Classification                                                                        & 30    \\ \hline
Probabilistic Reasoning                                                                      & 29    \\ \hline
Word Embeddings                                                                              & 25    \\ \hline
Hidden Markov Models                                                                         & 20    \\ \hline
NLP Resources                                                                                & 20    \\ \hline
Machine Translation Basics                                                                   & 19    \\ \hline
Monte Carlo Methods                                                                          & 19    \\ \hline
\end{tabular}
% \end{center}
    \caption{Counts of the most frequent taxonomy topic labels of the LectureBank files}
    \label{tab:stat_taxonomy}
\end{table}

\textbf{Vocabulary} Alongside our lecture files we also provide a vocabulary list containing 1,221 terms with the help of the LectureBank Corpus, the vocabulary can be used as an in-domain reference while it mainly focuses on NLP and related areas. Rather than using Wikipedia to create a corpus vocabulary, we took the union of three topic sets: the TutorialBank taxonomy topics, the 208 topics labeled for prerequisite chains and the topics extracted from LectureBank. The first two parts are provided by \cite{fabbri2018tutorialbank}, and we contributed more fine-grained topic words from our LectureBank. Different from \cite{fabbri2018tutorialbank} who manually propose topic terms, we propose topic terms in an automatical way: we found keywords from LectureBank by taking the header section of each individual lecture slide and post-processing and filtering that list. This method can be extended to other online resources such as blog posts and papers to enlarge the vocabulary. In the future, the vocabulary can be potentially used as additional concepts for creating a concept graph in an automated manner. This vocabulary can also serve an educational purpose for preparing lecture topics.

% Besides, an in-domain vocabulary as a dictionary is also essential for educational purposes such as an initial survey generation terms. 

\subsection{Comparison with Similar Datasets}

% TODO: remove

% \begin{table*}[h!]
% \small
%     \centering
% \begin{tabular}{ |c|c|c| }
% \hline
% Corpus & \#courses & \#domains \\ 
%  \hline
% MOOC's & 20 & 3 \\ 
% \hline 
% XuetangX, Coursera & 39 & 4\\ 
%  \hline
% LectureBank & 60 & 5\\ 
% \hline
% \end{tabular}
%     \caption{Comparison with Similar Datasets}
%     \label{tab:stat}
% \end{table*}

\textbf{MOOCs} In a recent research, \cite{pan2017prerequisite} evaluated on three corpora extracted from Massive Open Online Courses (MOOCs), containing three domains (Machine Learning, Data Structures and Algorithms and Calculus).  They are the first to study prerequisite relations in a MOOC corpus. The corpora contain video subtitles and speech scripts, and the number of topics ranges from 128 to 244. They used Wikipedia to help obtain entity representations. 

\textbf{Coursera and XuetangX} Similarly, \cite{pan2017course} constructed four course corpora in two domains (Computer Science and Economics) and in two languages (English and Chinese) from video captions. In each corpus, the number of courses varies from 5 to 18. Also, they have a  significant number of candidate concepts for each corpus which ranges from 27,571 to 79,009. Similarly, they also benefit from the help of Wikipedia\footnote{https://dumps.wikimedia.org/enwiki/20170120/} and the Baidu encyclopedia\footnote{https://baike.baidu.com/} when learning prerequisite relations. 

\par
\textbf{ACL Anthology Scientific Corpus} A corpus for prerequisite relations among topics was presented by\cite{gordon2016modeling}. They used topic modeling to generate a list of 305 topics from the ACL Anthology \cite{Bird:08}. While the focus of this corpus is NLP, the resources come from academic papers, while we focus on tutorials from TutorialBank and our corpus of lectures. Additionally, they only annotate a subset of the topics for prerequisite annotations while we annotate two orders of magnitude larger in terms of prerequisite edges and show strong inter-annotator agreement.

% moved to related works
% \textbf{TutorialBank} \cite{fabbri2018tutorialbank} introduce a manually-collected and categorized corpus of about 6,300  resources on NLP and the related fields of ML, AI, and IR, making it a complementary dataset to ours. This corpus emphasizes surveys, tutorials and resources other than scientific papers which are classified into a taxonomy of about 305 topics. This corpus was released with a list of 208 topics for which prerequisite relationships were annotated. However, only a single annotator annotated each relation and prerequisite annotations were ternary versus our binary classification. As there have not been any experiments on learning prerequisite chains using this corpus, we choose to take advantage of both TutorialBank and LectureBank to learn prerequisite relations. 

\section{Prerequisite Chain Learning}

In this section, we introduce our two-step framework: concept feature learning and a recently proposed neural graph-based method in addition to traditional classification methods. 
% For each step, we provide both traditional methods and deep-learning based methods. 

% Statistics
% \subsection{Topic Extraction}

% % building concept vocab
% % TODO why 25 here. just bc we had that many at the time? yes
% To select topics, we did text mining on a subset of our slides. We extracted all the title entries of each slide on about 500 slide files among 25 courses. In terms of PDF files, we use PDFMiner \footnote{https://github.com/euske/pdfminer} to extract the first line of each page. 
% %  TODO fix, vague. also I need to describe the ranking, how it's from the Gordon paper
% Then we applied some rules and re-ranked the extracted topics, and finally three people who have the domain knowledge did annotating on the selected topics on weather to keep a specific topic. Finally we have about 1000 meaningful topics extracted from the slide header entities. The top 10 topics are: \textit{nlp, markov models, pcfg, neural networks, viterbi algorithm, rnn, cky, perceptron, language modelling, n-gram}. 

\subsection{Concept Feature Learning}

% \subsubsection{Shallow Features with Topic Labeling}
% We applied LDA topic  modeling based on the TF-IDF features of each document. As a result, we can represent each topic or concept as a sparse vector along with a list of key words representing the topic. To match the extracted topic with our manually labeled prerequisite chain relations,  we did topic labeling for every topic. We followed the approach \cite{bhatia2016automatic} but instead of generating topic candidates from Wikipedia as a third-party corpus, we treat the topic extraction results from the last step as our topic candidates. 

% \subsubsection{Deep Features}

The first step is to extract concept vectors from various documents or lectures. We trained a Doc2Vec model \cite{le2014distributed}, an unsupervised model which can train dense vectors as representations for variable-length texts such as sentences or documents.  Our Doc2Vec model was trained using Gensim\footnote{https://radimrehurek.com/gensim/}. We set the dimension of the document vector to be 300, and the model was trained in the manner of Distributed Memory Model of Paragraph Vectors (PV-DM) \cite{le2014distributed}. We obtained document representations from our LectureBank data and TutorialBank data from \cite{fabbri2018tutorialbank} separately as well as after combining the corpora. We then took the trained Doc2Vec model to infer the embedded vector of a given concept. This dense vector topic representation is then used as input to our models.

\subsection{Prerequisite Chain Learning as Linking Prediction}

% \subsection{Learning Prerequisite Chains from Concept Features} 
% \subsection{Variational graph auto-encoders}
Concept prerequisite chain learning can be viewed as a type of link prediction problem with prerequisite relations as links among concepts. The goal of this model is to learn a directed graph $\mathcal{G} = (V, E)$, where the vertices $V$ correspond to concepts and the edges E correspond to the prerequisite relationships (if any) among the concepts. The model takes as input an adjacency matrix $A$ and a feature matrix $\textbf{X}$ of size $n \times d$, where $n$ is the number of vertices/concepts and $d$ is the number of input features per vertex.
\subsubsection{Graph Autoencoders}
In the case of the non-probabilistic GAE, a $n \times f$ embeddings matrix $\textbf{A}$, where $f$ is the size of the embeddings, is parameterized by a two-layer GCN: 
$$\textbf{Z} = \textit{GCN}(\textbf{X}, \textbf{A})$$
and the reconstructed adjacency matrix is: 
$$\hat{\textbf{A}} = \sigma(\textbf{ZZ}^T)$$
\subsubsection{Variational Graph Autoencoders}
As an extension to the above model, the  stochastic latent variables $z_i$, summarized in $\textbf{Z}$, are introduced  and $\textbf{Z}$ is modeled by a Gaussian prior distribution $\prod_i \mathcal{N}(\textbf{z}_i, \textbf{0}, \textbf{I})$.
As in \cite{kipf2016variational}, we use the following inference model parameterized by a two-layer GCN: 
$$q(\textbf{Z}|\textbf{X}, \textbf{A}) = \prod_{i=1}^{N} q(\textbf{z}_i| \textbf{X},\textbf{A}) $$
where
$$q(\textbf{z}_i|\textbf{X}, \textbf{A}) = \mathcal{N}(\textbf{z}_i|\boldsymbol{\mu}_i, \textit{diag}({\sigma}_i^2)),$$
the matrix of mean vectors is $\smash{\boldsymbol{\mu} = GCN_{\boldsymbol{\mu}}(\textbf{X},\textbf{A})}$  and $\smash{\log{\boldsymbol{\sigma}}=GCN_{\boldsymbol{\sigma}}(\textbf{X},\textbf{A})}$. The two-layer GCN is defined as:
$$GCN(\textbf{X},\textbf{A}) =  \widetilde{\textbf{A}}\textit{ReLU}(\widetilde{\textbf{A}}\textbf{X}\textbf{W}_0)\textbf{W}_1,$$
where $\textbf{W}_i$ represents the weight matrix at level $i$ and $\widetilde{\textbf{A}} = \textbf{D}^{-\frac{1}{2}}\textbf{A}\textbf{D}^{-\frac{1}{2}}$ is the symmetrically normalized adjacency matrix. 
The following generative model results:
$$P(\textbf{A}|\textbf{Z}) =  \prod_{i=1}^{N}\prod_{j=1}^{N}p(\textbf{A}_{ij} | \textbf{z}_i, \textbf{z}_j),$$
where
$$p(A_{ij} =1| \textbf{z}_i, \textbf{z}_j) = \sigma(\textbf{z}_i^T\textbf{z}_j)$$
The variational lower bound $\mathcal{L}$ is optimized \textit{w.r.t.} the variational parameters $\textbf{W}_i$:
% $$\mathcal{L} = E_{q(Z|X,A)}[\textit{log}p(A|Z)] -  KL[q(Z|X,A)||p(Z)],$$
\begin{align*}
\mathcal{L} &= E_{ q(\textbf{Z}|\textbf{X},\textbf{A}) }[ \log{p(\textbf{A}|\textbf{Z})}]- \\
& KL[q(\textbf{Z}|\textbf{X},\textbf{A})||p(\textbf{Z})] 
\end{align*}

where $KL[q(\cdot)||p(\cdot)]$ is the Kullback-Leibler divergence between $q$ and $p$.

\section{Experiments}
We treat the predictions of our prerequisite chain learning problem as a binary classification result among pairs of concepts and report precision, recall and F1 scores. We report results on 5 fold cross validation where the test set contains  $10\%$ of the positive prerequisite labels, following  \cite{kipf2016variational}. For the task of learning prerequisite chains, positive samples are usually rare, leading to imbalanced datasets. More specifically, we first divided training and testing where it was guaranteed that 10\% positive samples were selected in the testing set, then added the same number of negative samples into the testing set and took the rest samples as the training set. Finally, during each run, we oversampled on the training set, and report an average score of the 5 runs. We have 921 positive concept pairs and 41,829 negative concept pairs in total before oversampling.

% \subsection{Topic Vectors}

% \textbf{Topic Modelling} We apply the LDA model to construct shallow features. As a result, each topic is represented by a vector, with each dimension being a word from the vocabulary. In terms of the topic words, we take the top 50 words of each topic, then map them with our labelled topic collections to find the prerequisite relations. 

% \textbf{Clustering} Basic tf-idf model for comparison. 
% \textbf{Doc2Vec} We train our Doc2Vec model using Gensim \footnote{https://radimrehurek.com/gensim/index.html} library on the AAN resources dataset. We set the dimension to be 300, and the sliding window size is 10. 

%TODO: add how we adapt this method, i.e. how to calculate the weights. May delete
%  TODO fix this paragraph
\subsection{Models}
\textbf{Binary classifiers} We compared the neural graph-based methods with the following classifiers: Na\"ive Bayes classifier (\textit{NB}), SVM with linear kernel (\textit{SVM}), Logistic Regression (\textit{LR}) and Random Forest classifier (\textit{RF}). After obtaining the concept representations for each possible concept pair as described above, we concatenated the source concept and target concept embeddings together, used the corresponding prerequisite chain label as the class label and fit the classifiers.

\textbf{GAE} We used the same concept embeddings described above as vertex features for the GAE model. We followed the same parameters from \cite{kipf2016variational}. We trained using gradient descent on batches the size of the entire training dataset for 200 epochs using the Adam optimizer \cite{Kingma:14} and a learning rate of $0.01$. The two-layer GCN encoding contains 32 hidden units in the first layer and 16 hidden units in the second layer.

\subsection{Results and Analysis}

\begin{table}[t!]
\small
    \centering
\begin{tabular}{ |c  c c c| } 

\hline
 Method & Precision & Recall & F1\\ 

 \hline \hline 
 \textit{TutorialBank} & \multicolumn{3}{c|}{} \\ 
 
NB	&	0.761	&	0.453	&	0.567	\\
SVM	&	0.832	&	0.703	&	\textbf{0.761}	\\
LR	&	0.819	&	0.604	&	0.693	\\
RF	&	\textbf{0.871}	&	0.459	&	0.599	\\
GAE	&	0.634	&	0.884	&	0.725	\\
VGAE	&	0.599	&	\textbf{0.895}	&	0.717	\\

\hline

 \textit{LectureBank} &\multicolumn{3}{c|}{} \\ 
NB	&	\textbf{0.853}	&	0.611	&	0.710	\\
SVM	&	0.835	&	0.668	&	\textbf{0.740}	\\
LR	&	0.840	&	0.640	&	0.724	\\
RF	&	0.831	&	0.624	&	0.712	\\
GAE	&	0.577	&	0.905	&	0.705	\\
VGAE	&	0.545	&	\textbf{0.921}	&	0.684	\\

 \hline

 \multicolumn{2}{|c}{ \textit{TutorialBank + LectureBank}} & & \\ 
NB	&	0.614	&	0.670	&	0.641	\\
SVM	&	\textbf{0.824}	&	0.688	&	\textbf{0.748}	\\
LR	&	0.794	&	0.613	&	0.690	\\
RF	&	0.787	&	0.519	&	0.625	\\
GAE	&	0.594	&	0.899	&	0.715	\\
VGAE	&	0.578	&	\textbf{0.916}	&	0.708	\\

 \hline
\end{tabular}
    \caption{Experiment results using oversampling with Doc2Vec concept representations trained on three settings: TutorialBank, LectureBank and TutorialBank combined with LectureBank.}
    \label{tab:results_over}
\end{table}

\begin{figure*}[t!]
  \includegraphics[width=7in]{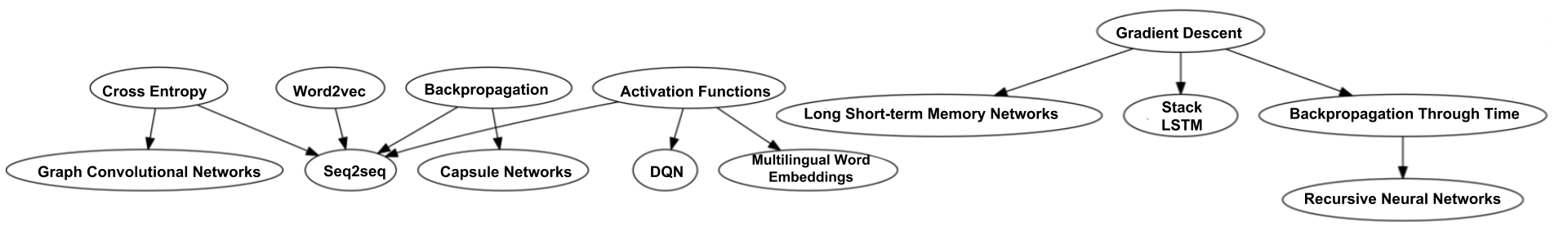}
  \caption[Caption for Prerequisites]{A subset of the recovered concept graph. The diagram is based on predictions on our test data. Some dependencies may be missing if the corresponding concept pairs were not in the test data or were predicted to have a negative label, e.g., the potential edge between \textit{Backpropagation Through Time} and \textit{Backpropagation}. } 
  \label{fig:recover}
\end{figure*}

% todo: move the gae and v-gae reference to GCN section
% description of result table
As shown in Table \ref{tab:results_over}, we report precision, recall and F1 scores of our embedding-based concept representation with the Na\"ive Bayes classifier, SVM with linear kernel, Logistic Regression and Random Forest classifier, along with the vanilla graph autoencoder (\textit{GAE}) and variational graph autoencoder (\textit{VGAE}). The concept representation was trained using Doc2Vec under three different settings: only using TutorialBank, only using LectureBank, and on the combination of TutorialBank + LectureBank.

% this is the analysis may need to rewrite this:
For the four binary classifiers, we observe a high precision and a low recall in all three different Doc2vec model settings. On the other hand, the GAE and VGAE show high recall and low precision. In general, SVM beats the other methods with high F1 scores among all corpora. The SVM and other basic classifiers benefit greatly from oversampling; we found that initial experiments with imbalanced datasets yielded very poor results. We modified the adjacency input to allow for parallel edges in a multi-graph to account for oversampled inputs in the GAE and VGAE, but the changes in performance were minimum. Intuitively, these models explicitly model the desired concept graph structure and are able to represent features in the vertices and propagate them through the graph. However, these specific networks were applied to the case of citation networks in which case parallel edges are non-existent. Thus we found that the SVM performed better in the general binary classification setting when using oversampling. 
\par
In terms of precision and recall for downstream tasks such as a search engine, for two classifiers with the same F1 score, we prefer the one with a higher recall. This coincides with our desired interface; we want the user to be able to mark subjects which are already known and be presented with potential prerequisites which are not known. We would rather give the student more concepts which they need to know rather than fewer in which case they may miss important fundamental knowledge. 

% corpora comparison
Although the highest F1 score was achieved when topic embeddings were trained on TutorialBank via SVM classifier, the variance between the three datasets is not significant. A potential explanation might be the that TutorialBank and LectureBank contain similar content and coverage. TutorialBank has notably four times as many documents and some notably longer resources such as topic surveys than LectureBank, which may improve the performance of our Doc2Vec document representation. However, we can see that the F1 score is only slightly higher than that of LectureBank. Additionally, the list of concepts was provided by \cite{fabbri2018tutorialbank} for the TutorialBank dataset, so we expected these topics to be broadly included in the TutoriaBank dataset.

% old results
% \begin{table}
% \small
%     \centering
% \begin{tabular}{ |c  c c c| } 

% \hline
%  Method & Precision & Recall & F1\\ 

%  \hline \hline 
%  \textit{TutorialBank} & \multicolumn{3}{c|}{} \\ 
 
%  NB	&	0.877	&	0.417	&	0.564	\\ 
%  RF & \textbf{0.960} &	0.315 &	0.471 \\
% %  RF	&	0.964	&	0.281	&	0.434	\\ 
% GAE	& 0.649 & 	\textbf{0.991} &	\textbf{0.784} \\
% V-GAE &	0.619	& 0.973 &	0.757 \\ \hline

%  \textit{LectureBank} &\multicolumn{3}{c|}{} \\ 
%  NB	&	0.913	&	0.485	&	0.633	\\ 
%  RF & \textbf{0.943} & 0.312	& 0.467 \\
%  GAE	& 0.616	& 0.967 &	\textbf{0.752}	\\
% VGAE &	0.540 &	\textbf{0.993} &	0.700 \\ \hline

%  \multicolumn{2}{|c}{ \textit{TutorialBank + LectureBank}} & & \\ 
%  NB	&	0.669	&	0.581	&	0.622	\\ 
%  RF & \textbf{0.956}	& 0.356	& 0.517 \\ 
%  GAE	& 0.596 & 	0.946 & 	0.731 \\
% V-GAE &	0.609 &	\textbf{0.954} & 	\textbf{0.744} \\
 
%  \hline

% \end{tabular}
%     \caption{Experiment Results without Oversampling}
%     \label{tab:results}
% \end{table}

\subsection{Concept Graph Recovery}

According to the F1 score, the highest-performing model is SVM with Doc2Vec trained on TutorialBank. To demonstrate the effectiveness of our model, we took a single training and testing fold from our 5-fold cross-validation experiments randomly and tried to recover the concept graph on the corresponding testing data by predicting the labels. Figure \ref{fig:recover} shows a subset of the recovered concept graph containing 14 vertices and 12 edges. We observe some reasonable paths, for example: \textit{Gradient Descent},  \textit{Backpropagation Through Time}, \textit{Recursive Neural Networks}. Another reasonable dependency relation shows that \textit{Seq2seq} is dependent upon multiple prerequisites such as \textit{Word2vec}, \textit{Backpropagation} and \textit{Activation Functions}.

\section{Conclusion and Future Work}

%contributions

In this paper, we introduced LectureBank, a collection of 1,352 English lecture slides from 60 university-level courses and their corresponding class label annotations. In addition, we extracted 1,221 concepts automatically from LectureBank which serve as additional references for in-domain vocabulary. We also release annotation of prerequisite relation pairs on 208 concepts. These annotations will be useful as both an educational tool for the NLP community and a corpus to promote research on learning prerequisite relations. 

In the future, we plan to expand LectureBank by enriching the coverage of courses, e.g., adding courses from medical information retrieval or linguistics, and we are planning to increase the corpus size to over 100 courses shortly. To take advantage of LectureBank and the prerequisite chains we have learned so far, we are planning to additionally classify each lecture into one of the prerequisite topics, thus making a search engine which can provide learning materials with the help of prerequisite relations. The search engine will first figure out the prerequisites of a query concept, and then it will be able to provide a list of lectures or even specific pages from them for each prerequisite concept, and in the future, it will take user experience as input to provide better resources for each specific user. Another part of future work is to automatically extract such topics. Finally, we plan to apply our prerequisite chain learning model to other applications such as reading list generation and survey generation.

% \section{Acknowledgments}

% \bigskip
% \noindent Thank you for reading these instructions carefully. We look forward to receiving your electronic files!

\bibliography{aaai}
\bibliographystyle{aaai}

\end{document}